# Dynamical behavior of passive particles with harmonic, viscous, and correlated Gaussian forces

Jae Won Jung,[1] Sung Kyu Seo,[1] and Kyungsik Kim [1,2,*]

[1] *DigiQuay Company Ltd., Seocho-gu, Seoul 06552, Republic of Korea*
[2] *Department of Physics, Pukyong National University, Busan 608-737, Republic of Korea*

In this paper, we study the Navier-Stokes equation and the Burgers equation for the dynamical motion of a passive particle with harmonic and viscous forces, subject to an exponentially correlated Gaussian force. As deriving the Fokker-Planck equation for the joint probability density of a passive particle, we find obviously the important solution of the joint probability density by using double Fourier transforms in three-time domains, and the moments from derived moment equation are numerically calculated. As a result, the dynamical motion of a passive particle with respect to the probability density having two variables of displacement and velocity in the short-time domain has a super-diffusive form, whereas the distribution in the long-time domain is obtained to be Gaussian by analyzing only from the velocity probability density. The moment $\mu_{2,2}$ particularly scales as $\sim t^5$ in the Navier-Stokes equation for $\tau = 0$ ($\tau$ = correlation time), and the moment in Burgers equation only with correlated Gaussian force reduces to $\mu_{2,2} \sim t^4$ in $t \gg \tau$ and $\tau = 0$.

DOI:

## I. INTRODUCTION.

Navier-Stokes equation is so far a problem of much interest in the various scientific systems. This equation in the full and simplified form has for a long time applied from the microscopic fluidity of the blood flow [1,2], turbulent simulation, convective diffusion contamination to macroscopic design of power station, the modelling of magneto-hydrodynamics, the design of aircraft and cars, and the other various systems. In addition, diverse turbulent models are studied the large eddy simulation [3,4], fluid-particle coupling method [5] including the cut-cell methods and the ghost-cell methods [6-9].

In 1961, Wyld [10] has concentrated on the theory of turbulence applying the developed formalism of Feynman diagrammatic techniques, as introducing the method of systematic perturbation similar to quantum field theory. Later, Lee [11] has generalized Wyld's formulation to the hydro-magnetic equations in the theory of the stationary, homogeneous, and isotropic turbulence in compressible fluids and to the solution for velocity and magnetic field in form of perturbation series. Forster *et al*. [12,13] have argued that in the infrared range the renormalization group theory can be applied and analyzed to dynamic behavior of velocity correlations generalized by Navier-Stokes equation in a randomly incompressible fluid. Such results have applied a forced Burgers equation in 1D, even though it is proved that the physical understanding of the Navier-Stokes equation below 2D is not clear. The Burgers equation has been first introduced by Batman and later studied by Burgers [14]. This equation is a convection-diffusion equation, as it is presently known, studied in various fields of the nonlinear acoustics, the traffic flow, the fluid turbulence, the gas dynamics, and the neural network [15-18], and so on. Navier-Stokes turbulence [19] also has shed light crucially on the 2D fluid for many years. Examples in 2D turbulences are the soap film flows [20,21], the magnetically stratified fluids [22], the two-fluid hydrodynamics [23,24], the plasmas in equatorial ionosphere [25], and the rotating fluids [26]. Over three decades ago, the fractional Brownian motion and the fractional diffusion motion have treated and analyzed in anomalous diffusions and transport processes theoretically and numerically [27-33].

Antonov *et al*. [34,35] have recently coupled to the Kazantsev-Kraichnan turbulent ensemble expressed to the environment of different models [36-38] such as the Kardar-Parisi-Zhang (KPZ) model [39-42], Hwa-Kardar model [43,44], Pavlik mode [45]. Their results have showed in an interesting and meaningful way the significant effect inducing non-linearity or making the anisotropic scaling through the renormalization group analysis.

Scientific researches for both the run-and-tumble particles and the self-propelled particles among active particle systems have considered the important issues estimating the dynamical behaviors of active particles, as solving and analyzing probability density functions in the stochastic processes and the transport processes. Examples of the run-and-tumble particle systems are bacterium [46], rotating flagella [47], and 1D [48-50], 2D [51-53] run-and-tumble particle systems. Several examples of robots in self-propelled particle systems are included vibro-robots [54,55], Hexbug crawlers [56], camphor surfers [57,58], and rotating robots [59].

*Contact author: kskim@pknu.ac.kr

We firstly derive the Fokker-Planck equation from the Navier-Stokes equation for a passive particle with the harmonic and viscous forces. We solve the joint probability density of the displacement and the velocity. Secondly, we simply introduce the Burgers equation for a passive particle. The probability density with the harmonic and viscous forces is only compared and analyzed to that with an exponentially correlated Gaussian force.

The paper is structured as follows: In section II, we derive the Fokker-Planck equation and briefly use the Fourier transform of the joint probability density of the Navier-Stokes turbulence equation in order to find the joint probability density and its mean squared values in three-time domains of correlation time. In section III, we clearly solve the joint probability density of the Burgers equation by using a similar method of section II. In section IV, we simply and easily calculate the kurtosis, the correlation coefficient, and the moment from the moment equation related to displacement and velocity in section IV. Lastly, we provide a conclusion summarizing our key findings in section V.

## II. NAVIER-STOHES EQUATION WITH HARMONIC AND VISCOUS FORCES

### A. Fokker-Planck equation

In this section, we derive Fokker-Planck equation from the equations of motion for a passive particle with harmonic and viscous forces expressed as follows:

$$\frac{d}{dt}v + \lambda v \cdot \nabla v = -\frac{1}{\rho}\nabla p + e\nabla^2 v - \beta x - rv + f(t). \quad (1)$$

Here the parameter $e$ denotes the drag coefficient, $-rv$ the viscous force, and $-\beta x$ a harmonic force acting on the particle. We put the pressure as $p = \rho v^2/2$, and $\lambda = 1$, $\rho = 1$ is the dimensionless value. The pressure terms in the above equations are not used to enforce the incompressibility condition $\nabla \cdot v = 0$. In Eq. (1), we add the random forces $f(t)$ by the fluctuation of the particle as follows:

$$<f(t)f(t')> = f_0^2 f(t-t') = \frac{f_0^2}{2\tau}\exp(-\frac{|t-t'|}{\tau}). \quad (2)$$

Here $f_0^2 = 2rk_B T_r$. The parameter $f_0$ denotes coupling strength, $T_r$ temperature, $k_B$ Boltzmann constant, and $\tau$ correlation time. Now the joint probability density $p(x(t),v(t),t)$ for the displacement $x$ and the velocity $v$ is defined by

$$p(x(t),v(t),t) = <\delta(x-x(t))\delta(v-v(t))>. \quad (3)$$

We assume from the joint probability density that the particle is initially at rest at time $t = 0$. By taking time derivatives in joint probability density and inserting Eq. (1) into Eq. (3), we can write the time derivative of the joint probability equation for $p(x(t),v(t),t) \equiv p$ and $\delta_x(x-x(t)) \equiv \delta_x$ as

$$\frac{\partial}{\partial t}p = -\frac{\partial}{\partial x}<\frac{\partial x}{\partial t}\delta(x-x(t))\delta(v-v(t))> -\frac{\partial}{\partial v}<[-v\nabla v -\frac{1}{2}\nabla v^2 + e\nabla^2 v - \beta x - rv + g(t)]\delta_x \delta_v>. \quad (4)$$

Then the joint probability density is derived as

$$\frac{\partial}{\partial t}p = -v\frac{\partial}{\partial x}p + \frac{\partial}{\partial v}v\frac{\partial}{\partial x}vp - \frac{1}{2}\frac{\partial^2}{\partial v \partial x}v^2 p + e\frac{\partial}{\partial v}\frac{\partial^2}{\partial x^2}vp + \beta x\frac{\partial}{\partial v}p + r\frac{d}{dv}vp - \alpha c(t)\frac{\partial^2}{\partial x \partial v}p + \alpha b(t)\frac{\partial^2}{\partial v^2}p. \quad (5)$$

Here the parameters are $\alpha = f_0^2/2$, $a(t) = 1 - \exp(-t/\tau)$ and $b(t) = (t+\tau)\exp(-t/\tau) - \tau$. Is clear that Eq. (5) is called the Fokker-Planck equation, as mentioned as Introduction. Some of the derivations in relation to the correlated Gaussian force are given in Eq. [60].

In order to find the joint probability density, we define $P(\xi,v,t)$, the double Fourier transform of the joint probability density, by the equation

$$P(\xi,v,t) = \int_{-\infty}^{+\infty}dx\int_{-\infty}^{+\infty}dv \exp(-i\xi x - ivv)P(x,v,t). \quad (6)$$

Thus the Fourier transform of Fokker-Planck equation, Eq. (5), is

$$\frac{\partial}{\partial t}P(\xi,v,t) = [-\beta v\frac{\partial}{\partial \xi} + [\xi - rv + \frac{\xi v}{2}\frac{\partial}{\partial v} - e\xi^2]\frac{\partial}{\partial v} + v\xi\frac{\partial^2}{\partial \xi \partial v}]P(\xi,v,t) + [\alpha b(t)\xi v - \alpha a(t)v^2]P(\xi,v,t). \quad (7)$$

### B. $P(x,t)$ and $P(v,t)$ in short-time domain

In this subsection, we will find the probability density $P(x,t)$ and $P(v,t)$ in the short-time domain $t << \tau$. In order to obtain the special solutions for $\xi$, $v$ by the variable separation from Eq. (7), the two equations of displacement and velocity are as follows:

$$\frac{\partial}{\partial t} P(\xi,t) = (-\beta v + \xi v D_v) \frac{\partial}{\partial \xi} P(\xi,t)$$
$$+ \frac{1}{2}\alpha[b(t)\xi v - a(t)v^2]P(\xi,t) + BP(\xi,t) \quad (8)$$

$$\frac{\partial}{\partial t} P(v,t) = (\xi - r_1 v + \xi v D_v / 2 - e\xi^2 v) \frac{\partial}{\partial v} P(v,t)$$
$$+ \frac{1}{2}\alpha[b(t)\xi v - a(t)v^2]P(v,t) - \frac{B}{2}P(v,t). \quad (9)$$

The constant $B$ denotes the separation constant. The differential symbol $D_\xi$ is given by $\partial/\partial\xi$. Presently, we assumes briefly from Eq. (9) that $[\beta v(1 - \xi D_v / \beta)]^{-1} \cong (\beta v)^{-1}(1 + \xi D_v / \beta)$. As we consider $\frac{\partial}{\partial t}P(\xi,t) = 0$ in the steady state and $P(\xi,t) \to P^{st}(\xi,t)$, we get Eq. (8) as

$$P^{st}(\xi,t) = \exp[\frac{\alpha}{2\beta v}[b(t)v\frac{\xi^2}{2} - a(t)v^2\xi] + \frac{\alpha D_v}{2\beta^2 v}[b(t)v\frac{\xi^3}{3}$$
$$- a(t)v^2\frac{\xi^2}{2}]P(v,t) + \frac{B}{2\beta v}(1 + \xi D_v / \beta)\xi]. \quad (10)$$

In order to solve the solution of the probability density for $\xi$ from $Q(\xi,t) \equiv R(\xi,t)Q^{st}(\xi,t)$, we continuously obtain the Fourier transforms of probability density via the calculation including terms up to order $t^2/\tau^2$, that is,

$$Q(\xi,t) = R(\xi,t)\exp[-\frac{\alpha}{2(\beta v)^2}[b'(t)v\frac{\xi^3}{6} - a'(t)v^2\frac{\xi^2}{2}]$$
$$- \frac{\alpha D_v}{2\beta^3 v^2}[b'(t)v\frac{\xi^4}{12} - a'(t)v^2\frac{\xi^3}{6}]] \quad (11)$$

$$R(\xi,t) = S(\xi,t)\exp[\frac{\alpha}{2(\beta v)^3}[b''(t)v\frac{\xi^4}{24} - a''(t)v^2\frac{\xi^3}{6}]$$
$$+ \frac{\alpha D_v}{2\beta^4 v^3}[b''(t)v\frac{\xi^4}{60} - a''(t)v^2\frac{\xi^4}{24}]]. \quad (12)$$

$$S(\xi,t) = T(\xi,t)\exp[-\frac{\alpha}{2(\beta v)^4}[b'''(t)v\frac{\xi^5}{120}]$$
$$- \frac{\alpha D_v}{2\beta^5 v^4}[b'''(t)v\frac{\xi^6}{360}]]. \quad (13)$$

Here $a'(t) \equiv d/dt$ and $a''(t) \equiv d^2/dt^2$. Discarding terms proportional to $1/\tau^3$ and taking the solutions as arbitrary functions of variable $t - \xi/(\beta v - \xi v D_v)$, the arbitrary function $T(\xi,t)$ becomes $\Theta[t - \xi/(\beta v - \xi v D_v)]$. As a result, we find that

$$P(\xi,t) = \Theta[t - \xi/(\beta v - \xi v D_v)]S^{st}(\xi,t)R^{st}(\xi,t)Q^{st}(\xi,t)P^{st}(\xi,t)$$
$$(14)$$

By the similar method (Eq. (10) – (14)) from Eq. (9) for $v$, we get the Fourier transform of probability density for the velocity as

$$P(v,t) = \Theta[t + v/(\xi - rv + \xi v D_v / 2 - e\xi^2 v)]$$
$$S^{st}(v,t)R^{st}(v,t)Q^{st}(v,t)P^{st}(v,t). \quad (15)$$

Therefore, from Eq. (14) and Eq. (15), we calculate the Fourier transforms of joint probability density as

$$P(\xi,t) = \exp[-\frac{3\alpha t^3}{4\beta\tau}\xi^2 - \frac{3\alpha t^3}{2}\xi v - \frac{\alpha t^3}{4\tau}v^2]$$
$$P(v,t) = \exp[-\frac{\alpha t^4}{6\tau}\xi^2 - \frac{2\alpha r^2 t^4}{\tau}v^2 - \frac{\alpha r t^4}{4\tau}\xi v]. \quad (16)$$

Using the inverse Fourier transform, we get

$$P(x,t) = [\pi \frac{3\alpha t^4}{\beta\tau}]^{-1/2}\exp[-\frac{\beta\tau x^2}{3\alpha t^4}] \quad (17)$$

$$P(v,t) = [8\pi \frac{\alpha r^2 t^4}{\tau}]^{-1/2}\exp[-\frac{\tau v^2}{8\alpha r^2 t^4}]. \quad (18)$$

The mean squared displacement and the mean squared displacement for $P(x,t)$ and $P(v,t)$ are, respectively, given by

$$<x^2> = \frac{3\alpha t^3}{2\beta\tau}, \quad <v^2> = \frac{4\alpha r^2 t^4}{\tau}. \quad (19)$$

### C $P(x,t)$ and $P(v,t)$ in long-time domain

Now we find $P(x,t)$ and $P(v,t)$ in long-time domain $t \gg \tau$. We write approximate equation for $\xi$ from Eq. (8) as

$$\frac{\partial}{\partial t}P_\xi(\xi,t) \cong \frac{\alpha}{2}[b(t)\xi v - \alpha_1 a(t)v^2]P_\xi(\xi,t). \quad (20)$$

In the steady state, the Fourier transform of probability density $P_\xi^{st}(\xi,t)$ is calculated as

$$P_\xi^{st}(\xi,t) = \exp[\frac{\alpha}{2}\int[b(t)\xi v - a(t)v^2]dt]. \quad (21)$$

We find the steady probability density $Q_\xi^{st}(\xi,t)$ for $\xi$ from $P_\xi(\xi,t) \equiv Q_\xi(\xi,t)P_\xi^{st}(\xi,t)$, that is,

$$Q_\xi^{st}(\xi,t) = \exp[\frac{\alpha}{2}\int[a(t)v^2 - b(t)\xi v]\,dt]. \quad (22)$$

From Eq. (8), the steady probability density $P^{st}(\xi,t)$ we obtained is not different as that of Eq. (10). As the Fourier transform of probability density $Q(\xi,t)$ in short-time domain is given by $Q(\xi,t) = R(\xi,t)Q_\xi^{st}(\xi,t)$, then $P(\xi,t)$ is derived as

$$P(\xi,t) = \Theta[t - \xi/(\beta v - \xi v D_v)]Q_\xi^{st}(\xi,t)P^{st}(\xi,t). \quad (23)$$

Applying Eq. (9) to the similar method from Eqs. (20)-(23) of $P(\xi,t)$ derived, we also get Fourier transforms of the probability density for velocity $v$ as

$$P(v,t) = \Theta[t + v/(\xi - rv + \xi v D_v / 2 - e\xi^2 v)]Q_v^{st}(v,t)P^{st}(v,t). \quad (24)$$

In the limit of the long-time domain, from Eq. (23) and Eq. (24), we have

$$P(\xi,v,t) = P(\xi,t)P(v,t)$$
$$= \exp[-\frac{\alpha t^4}{6}\xi^2 - \frac{\alpha r^2 t^4}{2}v\xi - \alpha r^2 t^3 v^2] \cdot \quad (25)$$

Using the inverse Fourier transform, the probability density $P(x,t)$ and $P(v,t)$ are, respectively, presented by

$$P(x,t) = [2\pi \frac{\alpha t^4}{3}]^{-1/2} \exp[-\frac{3x^2}{2\alpha t^4}] \quad (26)$$

$$P(v,t) = [4\pi \alpha r^2 t^3]^{-1/2} \exp[-\frac{x^2}{4\alpha r^2 t^3}] \cdot \quad (27)$$

The mean squared values $<x^2(t)>$ and $<v^2(t)>$ for the probability density $P(x,t)$ and $P(v,t)$ are, respectively, given by

$$<x^2(t)> = \frac{\alpha t^4}{3}, \quad <v^2(t)> = 2\alpha r^2 t^3 \cdot \quad (28)$$

### D. $P(x,t)$ and $P(v,t)$ in $\tau = 0$

In this subsection, we will find $P(x,t)$ and $P(v,t)$ in time domain $\tau = 0$. In time domain $\tau = 0$ ($a(t)=1$, $b(t)=0$), the approximate equation from Eq. (8) for $\xi$ is written as

$$\frac{\partial}{\partial t}P(\xi,t) \cong (-\beta v + \xi v D_v)\frac{\partial}{\partial \xi}P(\xi,t) - \frac{\alpha}{2}v^2 P(\xi,t) \cdot \quad (29)$$

In steady state, we can calculate $P^{st}(\xi,t)$ as

$$P^{st}(\xi,t) = \exp[-\frac{\alpha}{2\beta v}a(t)v^2\xi - \frac{\alpha D_v}{2\beta^2 v}a(t)v^2\frac{\xi^2}{2}] \cdot \quad (30)$$

We find the Fourier transform of probability density $P(\xi,t)$ as

$$P(\xi,t) = \Theta[t-\xi/(\beta v - \xi v D_v)]P^{st}(\xi,t) \cdot \quad (31)$$

By a similar method finding $P(\xi,t)$ for $\xi$, we get the Fourier transform of probability density

$$P(v,t) = \Theta[t+v/(\xi-rv+\xi v D_v/2 - e\xi^2 v)]P^{st}(v,t) \cdot \quad (32)$$

We can calculate $P(\xi,v,t)$ from Eq. (31) and Eq. (32) as follows:

$$P(\xi,v,t) = P(\xi,t)P(v,t)$$
$$= \exp[-\frac{\alpha r t^4}{8}\xi^2 - \frac{\alpha r t^3}{2}v\xi - \frac{\alpha t}{2}v^2] \cdot \quad (33)$$

Using the inverse Fourier transform, the probability density $P(x,t)$ and $P(v,t)$ are, respectively, presented by

$$P(x,t) = [\pi \frac{\alpha r t^4}{2}]^{-1/2} \exp[-\frac{2x^2}{\alpha r t^4}] \quad (34)$$

$$P(v,t) = [2\pi \alpha t]^{-1/2} \exp[-\frac{v^2}{2\alpha t}] \cdot \quad (35)$$

The mean squared displacement and the mean squared velocity are, respectively, given by

$$<x^2(t)> = \frac{\alpha r t^4}{4}, \quad <v^2(t)> = \alpha t \cdot \quad (36)$$

## III. BURGERS EQUATION WITH HARMONIC AND VISCOUS FORCES

### A. Fokker-Planck equation

In this subsection, we from now on study the Burgers equation for the dynamical motion of a passive particle with the harmonic and viscous forces, subject to an exponentially correlated Gaussian force. As deriving the Fokker-Planck equation for the joint probability density of a passive particle, we concentrate on solving and analyzing the solution of joint probability density by using the double Fourier transform in the limits of $t \ll \tau$, $t \gg \tau$ and for $\tau = 0$, where $\tau$ is the correlation time.

Firstly, we consider the Burgers equation with the correlated Gaussian force at 1D expressed as follows [12,14]:

$$\frac{d}{dt}v(t) + v(t) \cdot \nabla v(t) = \varepsilon \nabla^2 v(t) + g(t) \cdot \quad (37)$$

Here, $\varepsilon$ is the drag coefficient. One interests in ultra-violet behavior dominated by shock-wave excitation [61] with the restriction $\nabla \times v(t) = 0$, when we consider a velocity field of Eq. (37) at 3D. As is well known, an exponentially correlated Gaussian force depends on the time difference

$$<g(t)g(t')> = g_0^2 g(t-t') = \frac{1}{2\tau}\exp(-\frac{|t-t'|}{\tau}) \cdot$$
(38)

Here $g_0^2 = 2rk_B T$, the parameter $g_0$ denotes the coupling strength, $r$ viscous coefficient, $k_B$ Boltzmann constant, $T$ temperature, and $\tau$ the correlation time.

Secondly, we now introduce a modified Burgers equation with harmonic and viscous forces from Eq. (37) as

$$\frac{d}{dt}v + v \cdot \nabla v = \varepsilon \nabla^2 v - kx - rv + g(t) \cdot \quad (39)$$

Here $-kx$ denotes a harmonic force, $-rv$ viscous force, and the above equation is considered to be uneasily the novel equation of motion.

First of all, we briefly conscribe to the Burgers equation with harmonic and viscous forces. The joint probability density $P(x(t),v(t),t)$ for the displacement $x$ and the velocity $v$ is defined by

$$P(x(t),v(t),t) = <\delta(x-x(t))\delta(v-v(t))> \cdot \quad (40)$$

We assume that a passive particle is initially at rest at time $t = 0$. Taking time derivatives of the joint

probability density and inserting Eq. (39) into Eq. (40), we can write the equation [60] as

$$\frac{\partial}{\partial t}P(x,v,t) = -\frac{\partial}{\partial x}<\frac{\partial x}{\partial t}\delta(x-x(t))\delta(v-v(t))>$$
$$-\frac{\partial}{\partial v}<[-v\nabla v - rv - kx + \varepsilon\nabla^2 v$$
$$+g(t)]\delta(x-x(t))\delta(v-v(t))>. \quad (41)$$

Then the joint probability density is derived as

$$\frac{\partial}{\partial t}P(x,v,t) = [-v\frac{\partial}{\partial x} + \frac{d}{dv}v\frac{d}{dx}v - r\frac{d}{dv}v + kx\frac{\partial}{\partial v}]P(x,v,t)$$
$$+[\varepsilon\frac{d}{dv}\frac{d^2}{dx^2}v(t) - cb(t)\frac{\partial^2}{\partial x\partial v} + ca(t)\frac{\partial^2}{\partial v^2}]P(x,v,t) \quad (42)$$

Here $a(t) = 1 - \exp(-t/\tau)$, $b(t) = (t+\tau)\exp(-t/\tau) - \tau$ and $c = g_0^2/2$. It is clear that Eq. (42) is called the Fokker-Planck equation, as mentioned as Introduction. Some of the derivations in relation to the correlated Gaussian force are given in Ref. [60].

First of all, the double Fourier transform is defined by $P(\xi,v,t) = \int_{-\infty}^{+\infty}dx\int_{-\infty}^{+\infty}dv\exp(-i\xi x - ivv)P(x,v,t)$, in order to find the joint probability density analogously to section II. By taking the double Fourier transform, the Fourier transform of joint probability density as $a(t) \equiv a$ and $b(t) \equiv b$ is given by

$$\frac{\partial}{\partial t}P(\xi,v,t) = (\xi - rv - \varepsilon\xi^2 v + \xi v\frac{\partial}{\partial \xi})\frac{\partial}{\partial v}P(\xi,v,t)$$
$$-kv\frac{d}{d\xi}P(\xi,v,t) + c[b\xi v - av^2]P(\xi,v,t). \quad (43)$$

### B. $P(x,t)$ and $P(v,t)$ in short-time domain

In this subsection, we will find the probability density $P(x,t)$ and $P(v,t)$ in short-time domain $t \ll \tau$. Two equations, by variable separation for $\xi$, $v$, with respect to obtaining the special solutions are as follows:

$$\frac{\partial}{\partial t}P(\xi,t) = [-kv\frac{\partial}{\partial \xi} + \frac{c}{2}[b\xi v - av^2] + A]P(\xi,t) \quad (44)$$

$$\frac{\partial}{\partial t}P(v,t) = (\xi - rv - \varepsilon\xi^2 v + \xi v D_\xi)\frac{\partial}{\partial v}P(v,t)$$
$$+\frac{c}{2}[b\xi v - av^2]P(v,t) - AP(v,t). \quad (45)$$

Here $D_\xi \equiv \partial/\partial\xi$ and $A$ denotes the separation constant. As we have $\frac{\partial}{\partial t}P(\xi,t) = 0$ and $\frac{\partial}{\partial t}P(v,t) = 0$, the Fourier transform of the probability density $P(\xi,t)$ and $P(v,t)$ in the steady state becomes $P^{st}(\xi,t)$ and $P^{st}(v,t)$. Thus we have

$$[-kv\frac{d}{d\xi} + \frac{c}{2}[b\xi v - av^2] + A]P^{st}(\xi,t) = 0. \quad (46)$$

$$[(\xi - rv - \varepsilon\xi^2 v + \xi v D_\xi)\frac{\partial}{\partial v} + \frac{c}{2}[b\xi v - av^2] - A]P^{st}(v,t) = 0. \quad (47)$$

In the short-time domain $t \ll \tau$, Fourier transform of the probability density from Eq. (46) is obtained as

$$P^{st}(\xi,t) = \exp[\frac{c}{2kv}[\frac{b}{2}v\xi^2 - av^2\xi] + \frac{A}{kv}\xi]. \quad (48)$$

In order to find the Fourier transform of probability density for $\xi$ from $Q(\xi,t) \equiv R(\xi,t)Q^{st}(\xi,t)$, we obtain continuously distribution functions via the calculation including terms up to order $t^2/\tau^2$. That is,

$$Q(\xi,t) = R(\xi,t)Q^{st}(\xi,t)$$
$$= R(\xi,t)\exp[\frac{c}{2(kv)^2}[\frac{a'}{2}v^2\xi^2 - \frac{b'}{6}v\xi^3]] \quad (49)$$

$$R(\xi,t) = S(\xi,t)R^{st}(\xi,t)$$
$$= S(\xi,t)\exp[-\frac{c}{2(kv)^3}[\frac{a''}{6}v^2\xi^3 - \frac{b''}{24}v\xi^4]] \quad (50)$$

$$S(\xi,t) = T(\xi,t)S^{st}(\xi,t)$$
$$= T(\xi,t)\exp[\frac{c}{2(kv)^4}[\frac{a'''}{24}v^2\xi^4 - \frac{b'''}{120}v\xi^5]]. \quad (51)$$

Here $a' = \frac{da}{dt}$. Neglecting terms proportional to $1/\tau^3$ and taking the solution as an arbitrary function of the variable $t - (\xi/kv)$, arbitrary function $T(\xi,t)$ is given by $\Theta[t - (\xi/kv)]$. We consequently get $P(\xi,t)$ related to $T(\xi,t)$ as

$$P(\xi,t) = \Theta[t-(\xi/kv)]S^{st}(\xi,t)R^{st}(\xi,t)Q^{st}(\xi,t)P^{st}(\xi,t). \quad (52)$$

By expanding their derivatives to the second order in powers of $t/\tau$, we directly can obtain the Fourier transform of the probability density $P(\xi,t)$ after some cancellations, that is,

$$P(\xi,t) = \exp[-\frac{ct^2}{8k\tau^2}\xi^2 - \frac{ct^3}{4\tau}v\xi - \frac{ckt^3}{4}v^2]. \quad (53)$$

Here

$$\Theta[u] = \exp[-\frac{ck}{120\tau^2}v^2u^5 + \frac{ck}{48\tau}v^2u^4$$
$$+\frac{ckt}{12\tau}v^2u^3 - \frac{c}{4\tau}v^2u^2 - \frac{ckt}{4}v^2u^2]. \quad (54)$$

Using the inverse Fourier transform, the probability density $P(x,t)$ is presented by

$$P(x,t) = [\pi\frac{ct^2}{2k\tau^2}]^{-1/2}\exp[-\frac{2k\tau^2 x^2}{ct^3}]. \quad (55)$$

The mean squared displacement for $P(x,t)$ is given by

$$<x^2> = \frac{ct^2}{4k\tau^2}. \quad (56)$$

From Eq. (47) for $v$, the Fourier transform of the probability density in the steady state is obtained as

$$P^{st}(v,t) = \exp[\frac{c}{\xi}[\frac{a}{3}v^3 - \frac{b}{2}\xi v^2] + \frac{cr}{\xi^2}[\frac{a}{4}v^4 - \frac{b}{3}\xi v^3]$$
$$+ c\varepsilon[\frac{a}{4}v^4 - \frac{b}{3}\xi v^3] - \frac{ab}{\xi 3}v^3]. \quad (57)$$

When we derive $P^{st}(v,t)$ in case of the steady state of Eq. (47), we assume in first term of left hand side that $[\xi - rv - \varepsilon\xi^2 v + \xi v D_\xi]^{-1} \cong \frac{1}{\xi}[1 + rv\frac{1}{\xi} + \varepsilon\xi v - v D_\xi]$. Finding the solution of the joint function for $v$ from $Q(v,t) = R(v,t)Q^{st}(v,t)$, we continuously obtain the probability density in short-time domain $t \ll \tau$ via the calculation including terms up to order $t^2/\tau^2$. That is,

$$Q(v,t) = R(v,t)Q^{st}(v,t)$$
$$= \exp[\frac{c}{\xi^2}[\frac{a'}{12}v^4 - \frac{b'}{6}\xi v^3] + \frac{cr}{\xi^3}[\frac{a'}{20}v^5 - \frac{b'}{12}\xi v^4]$$
$$+ \frac{c\varepsilon}{\xi}[\frac{a'}{20}v^5 - \frac{b'}{12}\xi v^4] - \frac{c}{12\xi^2}b'v^4] \quad (58)$$

$$R(v,t) = S(v,t)R^{st}(v,t)$$
$$= \exp[\frac{c}{\xi^3}[\frac{a''}{60}v^5 - \frac{b''}{24}\xi v^4] + \frac{cr}{\xi^4}[\frac{a''}{120}v^6 - \frac{b''}{60}\xi v^5]$$
$$+ \frac{c\varepsilon}{\xi^2}[\frac{a''}{120}v^6 - \frac{b''}{60}\xi v^5] - \frac{c}{60\xi^3}b''v^5] \quad (59)$$

$$S(v,t) = T(v,t)S^{st}(v,t)$$
$$= \exp[-\frac{c}{\xi^3}\frac{b''}{120}v^5 - \frac{cr}{\xi^4}\frac{b''}{360}v^6 - \frac{c\varepsilon}{\xi^2}\frac{b''}{360}v^6 - \frac{cb'''}{360\xi^4}v^6] \quad (60)$$

Neglecting the terms proportional to $1/\tau^3$ and taking the solution as an arbitrary function of the variable $t + v/(\xi - rv - \varepsilon\xi^2 v + \xi v D_\xi)$, we put in first equality of Eq. (60) that an arbitrary probability density $T(v,t)$ is given by $\Theta[t + v/(\xi - rv - \varepsilon\xi^2 v + \xi v D_\xi)]$. We find that

$$P(v,t) = \Theta[t + v/(\xi - rv - \varepsilon\xi^2 v + \xi v D_\xi)]$$
$$\times S^{st}(v,t)R^{st}(v,t)Q^{st}(v,t)P^{st}(v,t). \quad (61)$$

By expanding their derivatives to second order in powers of $t/\tau$, we obtain the expression for $P(v,t)$ after some cancellations. That is,

$$P(v,t) = \exp[-\frac{ct^4}{8\tau}\xi^2 - \frac{crt^4}{4\tau}v\xi - \frac{2crt^4}{3\tau}v^2]. \quad (62)$$

Using the inverse Fourier transform, the probability density $P(v,t)$ is presented by

$$P(v,t) = [8\pi\frac{crt^4}{3\tau}]^{-1/2}\exp[-\frac{3\tau v^2}{8crt^4}]. \quad (63)$$

The mean squared velocity for $P(v,t)$ is given by

$$<v^2> = \frac{4crt^4}{3\tau}. \quad (64)$$

## C. $P(x,t)$ and $P(v,t)$ in long-time domain

In this subsection, we will find $P(x,t)$ and $P(v,t)$ in long-time domain $t \gg \tau$. In long-time domain, we write approximate equation for $\xi$, from Eq. (44),

$$\frac{\partial}{\partial t}P_\xi(\xi,t) \cong \frac{c}{2}[b\xi v - av^2]P_\xi(\xi,t). \quad (65)$$

Then we simply calculate $P_\xi(\xi,t)$ as

$$P_\xi(\xi,t) = \exp[\frac{c}{2}\int[b\xi v - av^2]\,dt. \quad (66)$$

Here $\int a(t)dt = t - \tau$, $a = 1$ and $\int b(t)dt = -\tau t$, $b = -\tau$ in long-time domain. We find the probability density $Q_\xi^{st}(\xi,t)$ for $\xi$ from $P_\xi(\xi,t) \equiv Q_\xi(\xi,t)P_\xi^{st}(\xi,t)$ as

$$Q_\xi^{st}(\xi,t) = \exp[\frac{c}{2}\int[av^2 - b\xi v]\,dt]. \quad (67)$$

From Eq. (10), we have the steady probability density $P^{st}(\xi,t)$ and the probability density $Q(\xi,t)$ in long-time domain $t \gg \tau$ as

$$P^{st}(\xi,t) = \exp[\frac{c}{2kv}[\frac{b}{2}v\xi^2 - av^2\xi]] \quad (68)$$

$$Q(\xi,t) = R(\xi,t)Q_\xi^{st}(\xi,t). \quad (69)$$

Taking the solutions as arbitrary functions of variable $t - \xi/kv$, we now select an arbitrary function $R(\xi,t)$ as $\Theta[t - (\xi/kv)]$. As expanding their derivatives to second order in powers of $t/\tau$, we obtain the expression for $P(\xi,t)$ after some cancellations, that is,

$$P(\xi,t) = R(\xi,t)Q_\xi^{st}(\xi,t)P^{st}(\xi,t)$$
$$= \Theta[t - [\xi/kv]]Q_\xi^{st}(\xi,t)P^{st}(\xi,t). \quad (70)$$

By using the method similar to the $\xi$'s case, we also write approximate equation for $v$ from Eq. (45) as

$$\frac{\partial}{\partial t}P_v(v,t) \cong c[b\xi v - av^2]P_v(v,t). \quad (71)$$

We simply calculate $P_v(v,t)$ as

$$P_v(v,t) = \exp[\frac{c}{2}\int[b\xi v - av^2]\,dt. \quad (72)$$

We find $Q_v^{st}(v,t)$ from $P(v,t) = Q_v^{st}(v,t)P_v(v,t)$ as

$$Q_v^{st}(\xi,t) = \exp[\frac{c}{2}\int[av^2 - b\xi v]\,dt]. \quad (73)$$

From Eq. (11), we have for long-time domain $t \gg \tau$

$$P^{st}(v,t) = \exp[\frac{c}{\xi}[\frac{a}{3}v^3 - \frac{b}{2}\xi v^2] + \frac{cr}{\xi^2}[\frac{a}{4}v^4 - \frac{b}{3}\xi v^3]$$
$$+ c\varepsilon[\frac{a}{4}\eta^4 - \frac{b}{3}\xi v^3] - \frac{c}{\xi}\frac{b}{3}v^3]] \quad (74)$$

$$Q_v(v,t) = R(v,t)Q_v^{st}(v,t). \quad (75)$$

As expanding their derivatives to second order in powers of $t/\tau$, we also obtain the expression for $P(v,t)$ after some cancellations. That is,

$$P(v,t) = R(v,t)Q_v^{st}(v,t)P^{st}(v,t) = \Theta[t + v/(\xi - rv - \varepsilon\xi^2 v + \xi v D_\xi)]Q_\eta^{st}(v,t)P^{st}(v,t). \quad (76)$$

For long-time domain, from Eq. (33) and Eq. (40), we have

$$P(\xi,v,t) = P(\xi,t)P(v,t)$$
$$= \exp[-\frac{ct^4}{6\tau}\xi^2 - \frac{cr^2t^4}{2}v\xi - \frac{cr^2t^3}{2}v^2]. \quad (77)$$

By using the inverse Fourier transform, the probability density $P(x,t)$ and $P(v,t)$ are, respectively, presented by

$$P(x,t) = [2\pi\frac{ct^4}{3\tau}]^{-1/2}\exp[-\frac{3\tau x^2}{2ct^4}] \quad (78)$$

$$P(v,t) = [2\pi cr^2t^3]^{-1/2}\exp[-\frac{x^2}{2cr^2t^3}]. \quad (79)$$

The mean squared values from Eq. (78) and Eq. (79) are, respectively, given by

$$<x^2(t)> = \frac{ct^4}{3\tau}, \quad <v^2(t)> = cr^2t^3. \quad (80)$$

### D. $P(x,t)$ and $P(v,t)$ in $\tau = 0$

In this subsection, we will find the probability density $P(x,t)$ and $P(v,t)$ in time domain $\tau = 0$ ($t \to \infty$). In $\tau = 0$ domain ($a=1, b=0$), we write the approximate equation from Eq. (44) for $\xi$

$$\frac{\partial}{\partial t}P(\xi,t) \cong -kv\frac{\partial}{\partial \xi}P(\xi,t) - \frac{c}{2}v^2P(\xi,t). \quad (81)$$

In steady state, we can calculate $P^{st}(\xi,t)$ as

$$P^{st}(\xi,t) = \exp[-\frac{c}{2kv}v^2\xi - \frac{A}{kv}\xi]. \quad (82)$$

We find the Fourier transform of probability density $P(\xi,t)$ as

$$P(\xi,t) = \Theta[t - \frac{\xi}{kv}]P^{st}(\xi,t) \quad (83)$$

From Eq. (45), we can find $P(v,t)$ by similar method calculating Eqs. (81) - (83). The Fourier transform of the joint probability density $P(\xi,v,t)$ is as follows:

$$P(\xi,v,t) = P(\xi,t)P(v,t) = \exp[-\frac{crt^4}{8}\xi^2 - \frac{crt^3}{2}v\xi - \frac{3cr^3t^4}{4}v^2]. \quad (84)$$

Using the inverse Fourier transform, the probability density $P(x,t)$ and $P(v,t)$ are, respectively, presented by

$$P(x,t) = [\pi\frac{crt^4}{2}]^{-1/2}\exp[-\frac{2x^2}{crt^4}] \quad (85)$$

$$P(v,t) = [\pi 3cr^3t^4]^{-1/2}\exp[-\frac{v^2}{3cr^3t^4}]. \quad (86)$$

The mean squared displacement and the mean squared velocity are, respectively, given by

$$<x^2(t)> = \frac{crt^4}{4}, \quad <v^2(t)> = \frac{3cr^3t^4}{2}. \quad (87)$$

## IV. MOMENT EQUATIONS

In this section, we derive the moment equation for $\mu_{m,n}$, i.e. $m$-th moment of displacement and $n$-th moment of velocity in the joint probability density. We also calculate the kurtosis and the correlation coefficient for displacement and velocity. Generally, the moment equation for Navier-Stokes equation with harmonic and viscous forces from Eq. (41) is given by

$$\frac{d\mu_{m,n}}{dt} = m\mu_{m-1,n+1} + \frac{1}{2}mn\mu_{m-1,n+1} - \varepsilon m(m-1)n\mu_{m-2,n} - \beta n\mu_{m+1,n-1}$$
$$- r\mu_{m,n} - mn\alpha b(t)\mu_{m-1,n-1} + n(n-1)\alpha a(t)\mu_{m,n-2}. \quad (88)$$

The moment equations for the Burgers equation with the correlated Gaussian force, harmonic and viscous forces from Eq. (42) are, respectively, derived as

$$\frac{d\mu_{m,n}}{dt} = -m\mu_{m-1,n+1} + mn\mu_{m-1,n+1} - \varepsilon m(n-1)\mu_{m-2,n}$$
$$- mncb(t)\mu_{m-1,n-1} + n(n-1)ca(t)\mu_{m,n-2}. \quad (89)$$

$$\frac{d\mu_{m,n}}{dt} = m\mu_{m-1,n+1} + mn\mu_{m-1,n+1} - r\mu_{m,n} - \varepsilon m(m-1)n\mu_{m-2,n}$$
$$- kn\mu_{m+1,n-1} - mncb(t)\mu_{m-1,n-1} + n(n-1)ca(t)\mu_{m,n-2}. \quad (90)$$

Here $\mu_{m,n} = \int_{-\infty}^{+\infty}dx\int_{-\infty}^{+\infty}dv x^m v^n P(x,v,t)$. In order to get the accuracy of Gaussian distribution, we use its higher moments as $\mu_{2m,0}(t) = (2m-1)!![\mu_{2,0}(t)]^m$. The kurtosis for displacement and velocity are, respectively, given by

$$K_x = <x^4>/3<x^2>^2 - 1, \quad K_v = <v^4>/3<v^2>^2 - 1. \quad (91)$$

We get the correlation coefficient as

$$\rho_{x,v} = <(x-<x>)(v-<v>)>/\sigma_x\sigma_v. \quad (92)$$

Here we assume that a passive particle is initially at $x = x_0$ and at $v = v_0$. $\sigma_x$ ($\sigma_v$) is the root-mean-squared displacement (velocity) of joint probability density.

TABLE I. Values of the kurtosis, the correlation coefficient and the moment $\mu_{2,2}$ in three-time domains,

for the Navier-Stokes equation with harmonic and viscous forces.

| Time | $x, v$ | $K_x, K_v$ | $\rho_{xv}$ | $\mu_{2,2}$ |
|---|---|---|---|---|
| $t \ll \tau$ | $x$ | $\dfrac{\beta\tau x_0^2}{\alpha}t^{-3}$ | $\dfrac{\sqrt{\beta}\tau x_0 v_0}{\alpha r}t^{-7/2}$ | $\dfrac{c\alpha^2}{(1+rt)\beta\tau}t^6$ |
|  | $v$ | $\dfrac{\beta\tau v_0^2}{\alpha r^2}t^{-4}$ | | |
| $t \gg \tau$ | $x$ | $\dfrac{x_0^2}{\alpha}t^{-4}$ | $\dfrac{\sqrt{\beta}\tau x_0 v_0}{\alpha r}t^{-7/2}$ | $\dfrac{c\alpha^2}{(1+rt)}t^6$ |
|  | $v$ | $\dfrac{v_0^2}{\alpha r^2}t^{-3}$ | | |
| $\tau = 0$ | $x$ | $\dfrac{x_0^2}{\alpha r}t^{-4}$ | $\dfrac{\sqrt{\beta}\tau x_0 v_0}{\alpha r}t^{-7/2}$ | $\dfrac{c\alpha^2 r}{(1+rt)}t^6$ |
|  | $v$ | $\dfrac{v_0^2}{\alpha}t^{-1}$ | | |

TABLE II. Values of kurtosis $K_x$ and $K_v$ in three-time domains for the Burgers equation with correlated Gaussian forces (CGF) and harmonic and viscous forces (HVF).

| Time | $x, v$ | CGF | HVF |
|---|---|---|---|
| $t \ll \tau$ | $x$ | $c^{-1}\tau x_0^2 t^{-4}$ | $c^{-1}k\tau^2 x_0^2 t^{-2}$ |
|  | $v$ | $c^{-1}\tau v_0^2 t^{-2}$ | $c^{-1}\tau v_0^2 t^{-4}$ |
| $t \ll \tau$ | $x$ | $c^{-1}x_0^4 t^{-3}$ | $c^{-1}r^{-1}\tau x_0^2 t^{-4}$ |
|  | $v$ | $c^{-1}v_0^2 t^{-1}$ | $c^{-1}r^{-2}v_0^2 t^{-4}$ |
| $\tau = 0$ | $x$ | $c^{-1}x_0^2 t^{-3}$ | $c^{-1}r^{-1}x_0^2 t^{-4}$ |
|  | $v$ | $c^{-1}v_0^2 t^{-1}$ | $c^{-1}r^{-3}v_0^2 t^{-4}$ |

TABLE III. Values of the correlation coefficient $\rho_{xv}$ in three-time domains for the Burgers equation with correlated Gaussian forces and harmonic and viscous forces.

| Time | $x, v$ | CGF | HVF |
|---|---|---|---|
| $t \ll \tau$ | $x, v$ | $c^{-1}\tau x_0 v_0 t^{-6}$ | $c^{-1}(r^{-1}\beta\tau^3)^{1/2}x_0 v_0 t^{-3}$ |
| $t \ll \tau$ | $x, v$ | $c^{-1}x_0 v_0 t^{-4}$ | $(cr)^{-1}\tau^{1/2}x_0 v_0 t^{-7/2}$ |
| $\tau = 0$ | $x, v$ | $c^{-1}x_0 v_0 t^{-4}$ | $c^{-1}r^{-2}x_0 v_0 t^{-4}$ |

TABLE IV. Values of the moment $\mu_{2,2}$ in three-time domains for the Burgers equation with the correlated Gaussian force and the harmonic and viscous forces.

| Time | $x, v$ | CGF | HVF |
|---|---|---|---|
| $t \ll \tau$ | $x, v$ | $\dfrac{a^2}{\tau^2}t^6$ | $\dfrac{a^2}{\tau(1+4\varepsilon t)}t^5$ |
| $t \ll \tau$ | $x, v$ | $\dfrac{a^2}{\tau^2}t^5$ | $\dfrac{a^2}{\tau(1+4\varepsilon t)}t^5$ |
| $\tau = 0$ | $x, v$ | $\dfrac{a^2}{\tau^2}t^5$ | $\dfrac{a^2}{\tau(1+4\varepsilon t)}t^5$ |

TABLE V. Comparison of the joint probability density, the mean squared displacement and the mean squared velocity for the Burgers equation with correlated Gaussian force and with harmonic and viscous forces. Here MSD, MSV, and JPD denotes the mean squared displacement, the mean squared velocity and the joint probability density, respectively.

| Time | Value | CGF | HVF |
|---|---|---|---|
| $t \ll \tau$ | JPD | $\exp[-\dfrac{4\tau x^2}{ct^4}]$ | $\exp[-\dfrac{2k\tau x^2}{ct^2}]$ |
|  | MSD | $\dfrac{ct^4}{8\tau}$ | $\dfrac{ct^2}{4k\tau}$ |
| $t \ll \tau$ | JPD | $\exp[-\dfrac{\tau v^2}{ct^2}]$ | $\exp[-\dfrac{2\tau v^2}{crt^4}]$ |
|  | MSV | $\dfrac{ct^2}{2\tau}$ | $\dfrac{crt^4}{2\tau}$ |
| $t \gg \tau$ | JPD | $\exp[-\dfrac{3x^2}{2ct^3}]$ | $\exp[-\dfrac{3\tau x^2}{2ct^4}]$ |
|  | MSD | $\dfrac{ct^3}{3}$ | $\dfrac{ct^4}{3\tau}$ |
| $t \gg \tau$ | JPD | $\exp[-\dfrac{v^2}{2ct}]$ | $\exp[-\dfrac{v^2}{2cr^2 t^3}]$ |
|  | MSV | $ct$ | $cr^2 t^3$ |
| $\tau = 0$ | JPD | $\exp[-\dfrac{3x^2}{ct^3}]$ | $\exp[-\dfrac{2x^2}{crt^4}]$ |
|  | MSD | $\dfrac{ct^3}{6}$ | $\dfrac{crt^4}{4}$ |
| $\tau = 0$ | JPD | $\exp[-\dfrac{v^2}{ct}]$ | $\exp[-\dfrac{v^2}{3cr^3 t^4}]$ |
|  | MSV | $\dfrac{ct}{2}$ | $\dfrac{3cr^3 t^4}{2}$ |

## V. CONCLUSIONS

In conclusion, we have investigated the Navier-Stokes equation and the Burgers equation for a passive particle with the harmonic and viscous forces, subject to an exponentially correlated Gaussian force. After deriving the Fokker-Planck equation for the joint probability density, we have solved approximately the solution of the joint probability density by using double Fourier transforms in three-time domains. As a result, the dynamical motion of particle with respect to the distribution of displacement and velocity in the short-time domain has a super-diffusive motion, whereas in the long-time domain the motion of particle is partially Gaussian by describing from the velocity probability density obtained.

In the Navier-Stokes equation, the mean squared displacement and the mean squared velocity for the joint probability density with harmonic and viscous forces of $x$ and $v$ is given by Eq. (19) in $t \ll \tau$ (Eq. (28) in $t \gg \tau$, and Eq. (36) in $\tau = 0$). The mean squared displacement (velocity) for a passive particle with harmonic and viscous forces behavior like the super-diffusion with $\sim t^3$ in $t \ll \tau$ ($t \gg \tau$), while mean squared velocity for a particle has the Gaussian form with time $<v^2> \sim t$ in $t \gg \tau$ and $\tau = 0$. The moment $\mu_{2,2}$ for a particle with harmonic and viscous forces scales as $\sim t^5$ in $\tau = 0$, consistent with our result. In Table I, we summarize calculated values for the kurtosis, the correlation coefficient, and the moment from moment equation.

In the Burgers equation, the mean squared velocity for a passive particle with harmonic and viscous forces behavior like the super-diffusion with $\sim t^4$ in $t \ll \tau$, $\tau = 0$, while the mean squared displacement for a passive particle with correlated Gaussian force to $\sim t^4$ in $t \gg \tau$, $\tau = 0$. Particularly, the mean squared velocity for a passive particle for a passive particle with harmonic and viscous forces has the ballistic form with time $<v^2> \sim t^2$ in $t \ll \tau$. Mean squared velocity for a particle with correlated Gaussian force has the Gaussian form, which is the normal diffusion with time $<v^2> \sim t$ in $t \gg \tau$. The moment $\mu_{2,2}$ for a particle with correlated Gaussian force scales as $\sim t^4$ in $t \gg \tau$, $\tau = 0$, consistent with our result. In Tables II, III, and IV, we summarize calculated values for the kurtosis, correlation coefficient, and moment from the moment equation in the Burgers equation for a passive particle with both correlated Gaussian force and harmonic and viscous forces. Table V is provided the comparison of joint probability density, mean squared displacement and mean squared velocity for the Burgers equation with both correlated Gaussian force and harmonic and viscous forces.

Despite much interest, the exact solutions for distributions of higher-order processes are rare, but the approximate solution of the probability distribution function is found in this paper. The approximate solution of the Navier-Stokes equation has not been solved yet, which is a simple approximate solution from our model. It is hoped in future that we extend our model to the generalized Langevin equation or the equation of motion with other forces. The results obtained can be compared and analyzed with other theories, computer-simulations, and experiments [62-68]. The other results will be continuously published in other journals.